\documentclass[manuscript]{aastex}

\shorttitle{The radio counterpart of HESS\,J1943+213}
\shortauthors{Gab\'anyi et al.}

\begin{document}


\title{VLBI search for the radio counterpart of HESS\,J1943+213}


\author{K. \'E. Gab\'anyi\altaffilmark{1}}
\affil{Konkoly Observatory, Research Centre for Astronomy and Earth
Sciences of the Hungarian Academy of Sciences, PO Box 67, Budapest,
H-1525, Hungary} \email{gabanyi@konkoly.hu}

\author{G. Dubner}
\affil{Instituto de Astronom\'{\i}a y F\'{\i}sica del Espacio
(CONICET-UBA), CC 67, Suc. 28, 1428, Buenos Aires, Argentina}

\author{E. Giacani\altaffilmark{2}}
\affil{Instituto de Astronom\'{\i}a y F\'{\i}sica del Espacio
(CONICET-UBA), CC 67, Suc. 28, 1428, Buenos Aires, Argentina}

\author{Z. Paragi}
\affil{Joint Institute for VLBI in Europe, Postbus 2, 7990 AA
Dwingeloo, The Netherlands}

\author{S. Frey}
\affil{F\"OMI Satellite Geodetic Observatory, PO Box 585, H-1592,
Budapest, Hungary}

\and

\author{Y. Pidopryhora}
\affil{Joint Institute for VLBI in Europe, Postbus 2, 7990 AA
Dwingeloo, The Netherlands}


\altaffiltext{1}{F\"OMI Satellite Geodetic Observatory, PO Box 585,
H-1592, Budapest, Hungary} \altaffiltext{2}{FADU, Universidad de
Buenos Aires, Argentina}


\begin{abstract}
HESS\,J1943+213, a TeV point source close to the Galactic plane
recently discovered by the H.E.S.S. collaboration, was proposed to
be an extreme BL Lacertae object, though a pulsar wind nebula (PWN)
nature could not be completely discarded. To investigate its nature,
we performed high-resolution radio observations with the European
Very Long Baseline Interferometry Network (EVN) and reanalyzed
archival continuum and H {\sc i} data. The EVN observations revealed a
compact radio counterpart of the TeV source. The low brightness
temperature and the resolved nature of the radio source are
indications against the beamed BL Lacertae hypothesis. The
radio/X-ray source appears immersed in a $\sim$ 1\arcmin~ elliptical
feature suggesting a possible galactic origin (PWN nature) for the HESS source. 
We found that HESS\,J1943+213 is located in the interior of a $\sim$1$\degr$ diameter
H {\sc i} feature, and explored the possibility of they being physically related.
\end{abstract}


\keywords{Radio continuum: general ---
Radio lines: ISM --- ISM: supernova remnants --- Techniques: interferometric --- X-rays:
individual: CXOU J194356.2+211823}



\section{INTRODUCTION}

The High Energy Stereoscopic System (H.E.S.S.) consists of four
imaging atmospheric Cherenkov telescopes situated in the Khomas
Highland of Namibia \citep{hess_define}. The H.E.S.S. Collaboration
has been surveying the Galactic plane for new very high energy
gamma-ray sources (VHE, that is with energies greater than 100 GeV).
Recently \cite{hess_disc} reported the discovery of an unresolved
VHE gamma-ray source close to the Galactic plane, HESS\,J1943+213,
with an integrated flux above 470 GeV corresponding to $\sim$ 2 per
cent of the Crab Nebula flux. Between 470 GeV and $\sim$ 6 TeV, the
spectrum of this source is well described by a power-law with photon
index $\Gamma = 3.1$.
 The conducted search for multi-wavelength counterparts revealed the presence of a hard X-ray source,
 observed by {\it INTEGRAL}, IGR J19443+2117, in the vicinity of the H.E.S.S. source. \cite{Landi} analyzed
 the data of the X-ray telescope on board the {\it Swift} satellite to search for soft X-ray counterparts
 of three {\it INTEGRAL} sources, among them IGR J19443+2117; they found a firm localization of this source
 in soft X-rays (SWIFT J1943.5+2120). Later, \cite{Tomsick} conducted {\it Chandra} X-ray observations of
 several {\it INTEGRAL} sources including IGR J19443+2117 to localize and measure their soft X-ray spectra.
 They concluded that IGR J19443+2117 is associated with the {\it Chandra } source CXOU J194356.2+211823 and
 the probability of spurious association is  only $0.39$\,per cent.
According to the precise {\it Chandra} measurement, the X-ray source
is located $23\arcsec$ away from HESS\,J1943+213 but still within
the H.E.S.S. error circle, leading \cite{hess_disc} to identify the
X-ray source with the H.E.S.S. source. All the proposed counterparts
of HESS\,J1943+213 discussed by \cite{hess_disc} are located within
the 68\,per cent best-fitting source position confidence level
contour of HESS\,J1943+213 \citep[see Fig. 6 in ][]{hess_disc}.

In a  search for a radio counterpart, a possible source is found in
the U.S. National Radio Astronomy Observatory (NRAO) Very Large
Array (VLA) Sky Survey (NVSS). The source, NVSS J194356+211826
\citep{nvss} is located $24\farcs7$ away from the H.E.S.S. position
(still within the H.E.S.S. error circle) and $3\farcs5$ away from
the {\it Chandra} position. The {\it Chandra} source is outside the
NVSS error circle.

\cite{hess_disc} discuss possible origins for HESS\,J1943+213,
proposing that it can be either a gamma-ray binary, a pulsar wind
nebula (PWN), or an active galactic nucleus (AGN). The binary
hypothesis is discarded because all known gamma-ray binaries contain
a massive bright stellar companion visible in the infrared/optical
band, and no evidence of such star is seen in the data, implying a
minimum distance of 25\,kpc, thus beyond the outer edge of the
Galaxy. The PWN hypothesis would be plausible, since the strong
magnetic fields cause rapid energy loss of accelerated particles,
preventing high-energy particles from moving far from the source,
like in the case of the Crab nebula, making  this kind of sources to
appear point-like on the scale of the H.E.S.S. angular resolution. In
fact, thirty-four PWNe have been detected up to date in the TeV
range\footnote{http://tevcat.uchicago.edu/ TeVCat is maintained by
S. Wakely and D. Horan, and is partially supported by the NSF.}, all
of them point-like in appearance, constituting the most abundant
class of VHE sources detected by Cherenkov telescopes. In the case
of HESS\,J1943+213, \cite{hess_disc} conclude that the small
gamma-ray to X-ray flux ratio implies that, if this is the origin of
the VHE emission, the nebula must be very young, approximately one
thousand years old, and located at a distance of $\le$16\,kpc.
However, the VHE spectrum of HESS\,J1943+213 is significantly softer
than the spectra of all known VHE PWNe, a fact that, when considered
together with the lack of extended X-ray emission in the Chandra
images, weaken the hypothesis of the VHE source being a PWN.
Finally, \cite{hess_disc} discuss the possibility of HESS\,J1943+213
being a blazar. Blazars are AGNs where a relativistic jet pointing
close to the line of sight produces Doppler-boosted emission
\citep{uniAGN}. According to \cite{hess_disc}, HESS\,J1943+213 would
be classified as a radio-loud, X-ray-strong BL Lac object belonging
to the high-frequency-peaked BL Lac (HBL) class. The drawback of this hypothesis is
the lack of high-energy counterpart (in the 100 MeV--100 GeV domain), a characteristic that
is unusual for these objects. Also the clearest characteristic of
blazars, their variability on all timescales, was not detected in
the case of HESS\,J1943+213.

After weighting the different pros and cons, \cite{hess_disc}
conclude that the identification of HESS\,J1943+213 as an extreme
HBL object is the most plausible origin.

Since the PWN nature cannot be completely rejected, it is very
important to identify as accurately as possible the counterparts of
HESS\,J1943+213 in other spectral ranges with the purpose of
understanding the true nature of the VHE emission. With this
objective in mind, we conducted exploratory Very Long Baseline
Interferometry (VLBI) continuum observation of NVSS J194356+211826
(hereafter J1943+2218) with the European VLBI Network (EVN) at
1.6~GHz frequency. This observation allowed us to spatially resolve
the radio source and to obtain its positional information with the
precision of a few milli-arcseconds (mas). We complemented the high
angular resolution study with the analysis of archival 1.4-GHz VLA
radio continuum data at intermediate angular resolution, and
investigated the Galactic interstellar gas in an extended field
around the VHE source. We describe the observations and data
reduction in Sect.~\ref{obs}, and present our results in
Sect.~\ref{res}. In Sect.~\ref{disc}, we discuss our findings, in
the light of possible scenarios: the source being HBL
(Sect.~\ref{HBL}), a gamma-ray binary (Sect.~\ref{gamma}) or a PWN
(Sect.~\ref{PWN}). We summarize our work in Sect.~\ref{concl}.


\section{OBSERVATIONS AND DATA REDUCTION}
\label{obs}


The exploratory EVN observation of J1943+2118 took place on 2011 May
18 (project id: RSG03). At a recording rate of up to 1024~Mbit~s$^{-1}$, seven antennas
participated in this e-VLBI experiment: Effelsberg (Germany),
Jodrell Bank Lovell Telescope (UK), Medicina (Italy), Onsala
(Sweden), Toru\'n (Poland), Hartebeesthoek (South Africa), and the
phased array of the Westerbork Synthesis Radio Telescope (WSRT, The
Netherlands). In an e-VLBI experiment \citep{Szomoru08}, the signals
received at the remote radio telescopes are streamed over optical
fiber networks directly to the central data processor for real-time
correlation. The correlation took place at the EVN data processor in
the Joint Institute for VLBI in Europe (JIVE), Dwingeloo, The
Netherlands, with 2~s integration time. The observations lasted for
2 hours. Eight intermediate frequency channels (IFs) were used in
both right and left circular polarizations. The total bandwidth was
128 MHz per polarization.

The target source was observed in phase-reference mode to obtain
precise relative positional information. It was crucial for
strengthening the identification of the source, since there was a
significant difference between the NVSS peak position and the {\it
Chandra} position. The phase-reference calibrator source J1946+2300
is separated from the NVSS radio source by $1\fdg77$ in the sky. Its
coordinates in the current 2nd realization of the International
Celestial Reference Frame (ICRF2) are right ascension
$\alpha_0=19^\mathrm{h}46^\mathrm{m} 6\fs25140484$ and declination
$\delta_0=+23\degr 0\arcmin 4\farcs4144890$ \citep{Fey09}. The
target--reference cycles of $\sim$5\,min allowed us to spend
$\sim$3.5\,min on the target source in each cycle, thus providing
almost 1.3\,h total integration time on J1943+2118. The source
position turned out to be offset from the phase center position
(taken from the NVSS catalog) by $\sim$$4\arcsec$, but still within
the undistorted field of view of the EVN.

\begin{figure}
   \centering
   \includegraphics*[width=8cm]{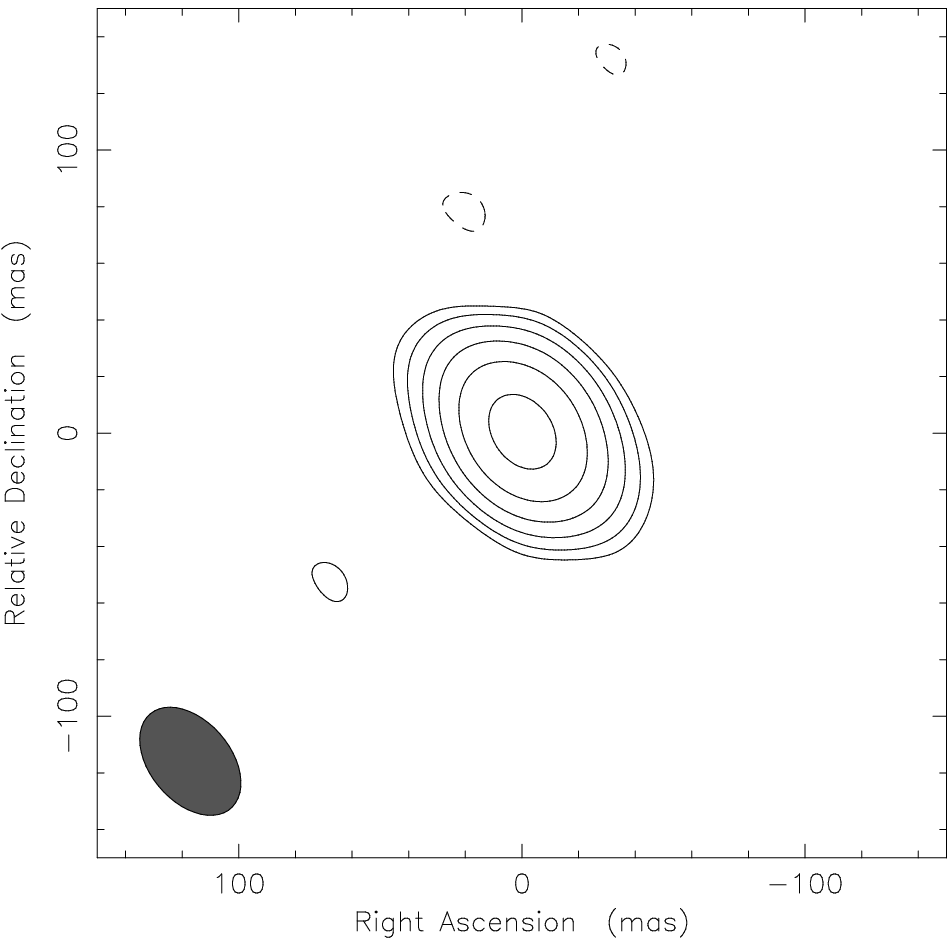} 
      \caption{1.6-GHz EVN image of J1943+2118. The lowest contours are drawn at $\pm0.6$\,mJy/beam. The positive contour levels increase by a factor of 2. The peak brightness is $25.3$\,mJy/beam. The Gaussian restoring beam shown in the lower-left corner is $43.9$\,mas $\times$ $28.5$\,mas (full width at half maximum, FWHM) with major axis position angle $40\fdg3$.}
         \label{ourEVN}
   \end{figure}

The NRAO Astronomical Image Processing System \citep[AIPS,
e.g.][]{aips} was used for the data calibration in the standard way.
We refer to \cite{frey2008} for the details of the data reduction
and imaging. The calibrated data were exported to the Caltech Difmap
package \citep{difmap} for imaging. Phase self-calibration was only
performed at the most sensitive antennas (Effelsberg, Jodrell Bank,
WSRT). No amplitude self-calibration was applied. Finally, the
longest baselines (from European telescopes to Hartebeesthoek) were
excluded from the imaging because the signal was barely above the
noise level due to the resolved nature of the source. The resulting
image of J1943+2118 is displayed in Fig.~\ref{ourEVN}.

We also analyzed archival VLA continuum data (project AH196,
observing date 1985 September 30). These observations were taken
from the region around our target source at 1.4~GHz, in the C
configuration of the array, thus improving the angular resolution by
a factor of three compared to the NVSS survey, which was performed
with the most compact D configuration of the VLA.

\begin{figure}
   \centering
   \includegraphics*[width=8cm]{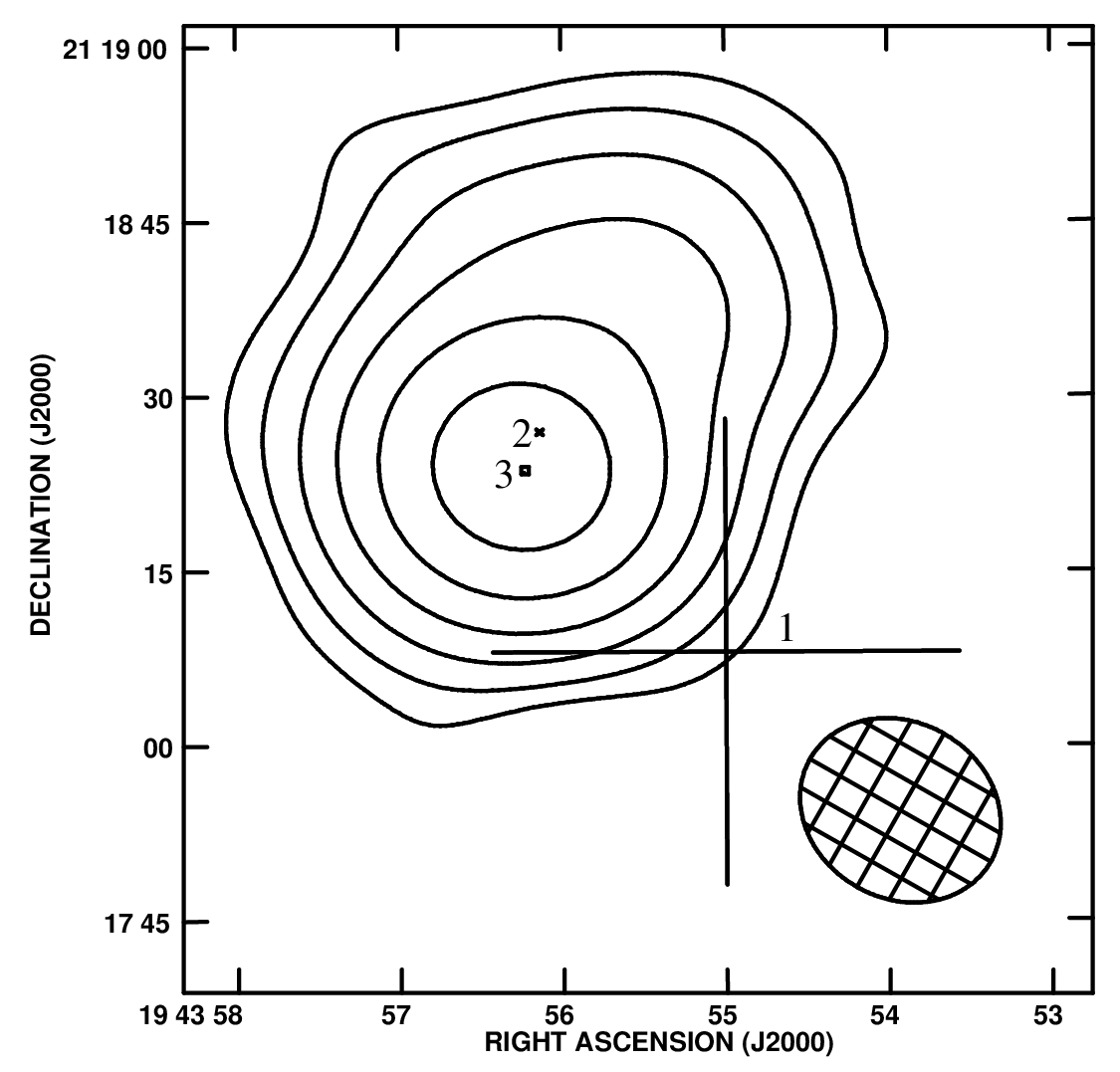} 
   \caption{1.4-GHz VLA C-configuration image of J1943+2118. The peak brightness is $60$\,mJy/beam. The lowest contours are drawn at $\pm 1.2$\,mJy/beam and the positive contour levels increase by a factor of 2. The Gaussian restoring beam shown in the lower-right corner is $17\farcs8 \times 15\farcs1$ with major axis position angle $60\fdg1$. Label number 1 indicates the TeV source position and its 90\,per cent error measured by H.E.S.S. The NVSS source position and the position of the X-ray counterpart is labeled as number 2 and number 3, respectively. The sizes of the symbols represent the errors of the corresponding position measurements. The {\it Chandra} position coincides with the position of the radio source derived from our phase-referenced EVN observation. (The accuracy of the EVN position is superior to that of the X-ray, therefore it cannot be distinctively displayed in this figure.)}
         \label{archVLA}
   \end{figure}

The archival experiment AH196 included five different pointings
covering a region of about $2\fdg4$$\times$$1\degr$ around our
source of interest. The calibration was performed in AIPS, using
3C~48 as the primary flux density calibrator. We used Difmap to
obtain the image shown in Fig.~\ref{archVLA}.

We also investigated the neutral hydrogen (H {\sc i}) radio emission
in a large field around the VHE source. To carry out this study, we
made use of the data from the VLA Galactic Plane Survey
\citep[VGPS,][]{VLAGPS}.


\section{RESULTS}
\label{res}

\subsection{Radio Continuum}
\subsubsection{EVN Results} \label{EVNres}

The phase-referenced exploratory EVN observation provided accurate
equatorial coordinates for J1943+2118: $\alpha=19^\mathrm{h}
43^\mathrm{m} 56\fs2372 \pm 0\fs0001$ and $\delta=21\degr 18\arcmin
23\farcs402 \pm 0\farcs002$. This position agrees well, within the
{\it Chandra} uncertainties, with the coordinates of
CXOU\,J194356.2+211823, the X-ray source proposed to be the
counterpart of the H.E.S.S. point source \citep{hess_disc}.
Therefore we confirm that they are the same object which is very
likely associated with the VHE emission detected by H.E.S.S. It has
to be noted that the difference between the new, accurate position
measurement and the position given in the NVSS catalog is
$3\farcs75$.

We used the Difmap package to fit a circular Gaussian brightness
distribution model component to the VLBI visibility data. The
feature in our EVN image can be well described with a component of
31\,mJy flux density and 15.8\,mas angular size (full width at half
maximum, FWHM).
Since J1943+2118 is very close to the Galactic plane (at about
$-1\fdg3$ Galactic latitude), angular broadening caused by the
intervening ionized interstellar matter can distort the image of a
distant compact radio source. It is important to disentangle
whether the observed size of the source is intrinsic or it is an
observational effect. According to the model of \cite{ne2001}, the
maximal amount of angular broadening of a point source in this
direction at 1.6\,GHz (the frequency of our EVN observation) is
expected to be 3.34\,mas. Therefore, even if we take this
effect into account, the ``de-broadened'' source size remains quite
large.

The flux density recovered in our high-resolution EVN observation is
only one-third of the value reported by NVSS at 1.4~GHz,
(102.6$\pm$3.6)\,mJy. To investigate the discrepancy between the
flux density values, we analyzed the WSRT synthesis array data taken
during our EVN experiment. The obtained flux density value,
$\sim$$95$\,mJy, still agrees well with the value reported in the
NVSS for J1943+2118. Thus, we can conclude that J1943+2118 is
likely to be extended and the high-resolution EVN observation
resolved out a significant portion of its large-scale
structure. This conjecture is confirmed with the archival VLA
C-array data (Section 3.1.2,  Fig. \ref{archVLA}). According to
those, the flux density of the source is (91$\pm$5)\,mJy, which
agrees with the flux density values derived from the two other (the
NVSS and the WSRT-only) lower-resolution data sets.

\subsubsection{VLA C-Array Results}

The $\sim$16\arcsec ~FWHM resolution image obtained from the VLA C-array
archival observations revealed an elongated structure with a size of
1\farcm 1 $\times$ 0\farcm 8~ (Fig. \ref{archVLA}). This shape can
naturally explain why the NVSS cataloged position, obtained from
lower resolution observations, is $\sim 4\arcsec$ off the position
of the EVN point source. In Fig. \ref{archVLA}, label 1 indicates
the TeV source position (and the 90 per cent error box), label 2
indicates the NVSS tabulated position, and label 3 indicates the
coinciding {\it Chandra} X-ray and EVN radio positions. The shift
between points 2 and 3 is exactly in the direction of the source
extension.

\subsubsection{Radio Continuum Emission at Large Scale}

Finally, to gain insight into the field where HESS\,J1943+213 is
located, and to search for possibly associated extended emission around the gamma-ray source, we
inspected the VGPS image of the radio continuum emission at 1.4~GHz
in a $5\degr$$\times$$5\degr$ area around the source position. 
The field appears almost empty around J1943+2118
with no trace of diffuse emission at the sensitivity of this survey.

\subsection{H {\sc i} Emission around the Position of HESS\,J1943+213}

\begin{figure}
   \centering
   \includegraphics*[width=8cm]{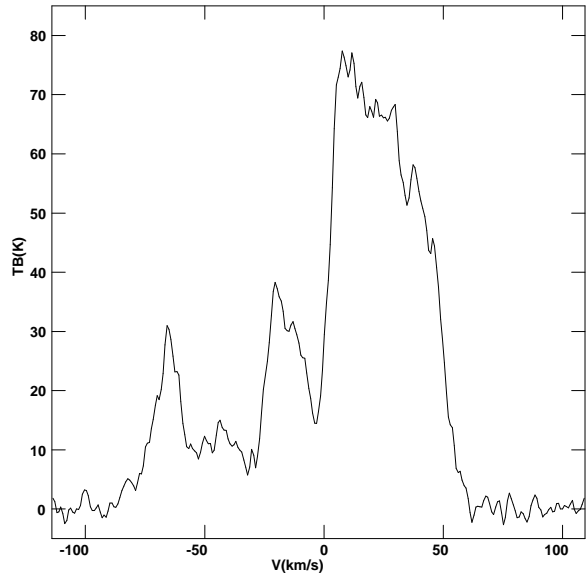} 
      \caption{H {\sc i} spectrum towards J1943+2118. The line of sight along l$= 57\degr.7$ in the direction of this source, crosses the spiral arms Perseus, Scutum and Sagittarius between the LSR velocities 0 and +60 {km\,s$^{-1}$}, in the interior of the Solar circle, and again the Cygnus arm around $-20$ {km\,s$^{-1}$} and the Scutum-Crux arm near $-65$ {km\,s$^{-1}$} corresponding to distances of about 10-11 and 15-16~\,kpc, respectively, in the outer Galaxy.}
         \label{HIspectrum}
   \end{figure}


\begin{figure}
   \centering
    \includegraphics*[width=8cm]{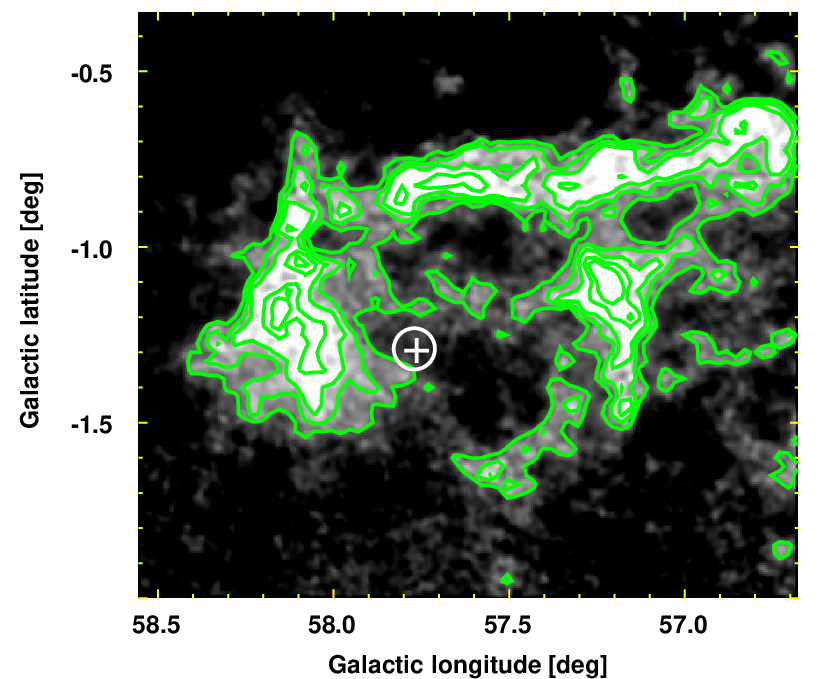} 
      \caption{A detailed view of the H {\sc i} feature surrounding the location of HESS\,1J943+213. The cross shows the location of J1943+2118 and the circle is the $2\farcm8$ confidence size of HESS\,J1943+213.}
         \label{detailedHI}
   \end{figure}

We also analyzed the H {\sc i} emission in a large
$5\degr$$\times$$5\degr$ field around the source, across the whole
observed velocity range between $-114$\,{km\,s$^{-1}$} and
+166\,{km\,s$^{-1}$}, using data extracted from the VGPS
\citep{VLAGPS}. We found that in all channel maps between radial
velocity $v$$\approx$$+50$\,{km\,s$^{-1}$} and
$v$$\approx$$+57$\,{km\,s$^{-1}$}, there is a striking, almost
complete shell-like feature surrounding the location of J1943+2118
(Fig.~\ref{detailedHI}). This H {\sc i} feature appears at a ``forbidden'' velocity
for the expected rotational properties of the Galactic gaseous disk,
since in this direction of the Galaxy the maximum positive velocity
(corresponding to the tangent point) is predicted to be about
$+35$\,{km\,s$^{-1}$}.
We also note that in this direction of the Galaxy, approximately in
middle of the first quadrant, the Galactic disk warps towards
positive latitudes \citep{voskes_thesis}, therefore it is very
likely that a feature like this shell, located at a forbidden
velocity and at a negative galactic latitude, is unrelated with the
normal Galactic gas.

Long time ago \cite{archival_mention}, working on a kinematical
analysis of  a large number of H {\sc i} profiles around $l=60\degr$,  
$b=0\degr$, reported the discovery of a large
ring-like object remarkable because its radial velocity (between
approximately +46 and +50 km\,s$^{-1}$) exceeded the maximum rotation
curve velocity. \cite{archival_mention} suggested that the object would be  the remnant
of a large-scale explosive event that took place near the Galactic
plane that at present  had been almost completely decelerated. The H {\sc i} feature
that we identified  around HESS\,J1943+213 lies in the periphery of that object 
and it may or may not be part of the same event (in fact
\cite{archival_mention} kinematically identifies the same feature, but assumes that
it is unrelated because its systemic velocity is larger than +48
km\,s$^{-1}$, a rather  arbitrary limit chosen to separate
components). In any case, this precursory study called the attention
towards the possible existence of explosive and/or expansive events 
that took  place in
this region of the Galaxy in the past. Several of such large H {\sc i} shells have been 
detected  in our Galaxy  \citep[e.g.][]{mcclure2002} and their origin 
is usually associated with the combined action of stellar winds and  subsequent 
supernova explosions inside the previously evacuated cavity. It has been proposed 
 that the expansion of such shells can lead to the sequential 
formation of new stars \citep[e.g.][]{starform,supershell}, and the most massive of them can end their lives exploding as supernovae.
As the radio emission associated with HESS\,J1943+213 is not strictly point-like but, on the contrary, there  is evidence that it is spatially extended (with an angular size of about $1\arcmin$ according to the VLA data), we explored the hypothesis that the gamma-ray emission originates in a PWN formed after the explosion of one of the second generation stars. H {\sc i} shells are usually  the last vestiges of explosions whose other manifestations already vanished.

The detected H {\sc i} shell is centered at $\alpha=19^\mathrm{h} 43^\mathrm{m} 23^\mathrm{s}, \delta=+21\degr 10\arcmin 45\arcsec  ({\it l}=57\fdg6, {\it b}=-1\fdg25 ) $, has an angular diameter $\sim  1^\circ$, and it can be clearly seen between + 50 and +57  km\,s$^{-1}$. Its anomalous velocities preclude the application of Galactic circular rotation models to independently estimate its distance, but if we assume that the detected H {\sc i} shell is physically associated with HESS\,J1943+213, it has to be at the same distance. 

Very recently on the basis of an H {\sc i} absorption spectrum
towards J1943+2118, \citet{leahy2012} proposed that it is
located beyond 16 kpc from the Sun. They considered this as an
evidence of HESS\,J1943+213 being an extragalactic source.  We note
that, according to the recent velocimetry and cartography of the
Milky Way presented by \citet{vallee2008}, in this direction of the
Galaxy (Galactic coordinates of HESS\,J1943+213: $l$=$57.7\degr$,
$b$=$-1.29\degr$), the farthest Galactic arm reaches distances
greater than 20 kpc. Following the best up-to-date velocimetric
models \citep{vallee2008}, the line of sight along $l$=$57\degr$
traverses five spiral arms: Sagittarius, Scutum and Perseus within
the Solar circle, and Cygnus, Scutum-Crux and again Sagittarius in
the outer Galaxy, up to at least 22 kpc from the Sun.
Fig.~\ref{HIspectrum} displays a H {\sc i} profile traced towards
the point radio source. Here it is possible to trace the presence of
H {\sc i } gas above 3$\sigma$ noise level at least up to
v$_{\mathrm {LSR}} \sim -82 $ {km\,s$^{-1}$}, corresponding to a
distance of $\sim$ 18 kpc according to the standard Galactic
circular rotation model \citep*{galrot}. Such a distance is consistent with the column density obtained from dust maps, and the $N_{\rm H}$ obtained from {\it Chandra} and {\it Swift} X-ray spectra, as presented by \citet{hess_disc}.
If we assume as a working hypothesis that the observed H {\sc i} shell is
related to the detected radio source and hence to
HESS\,J1943+213, an approximate distance of 17 kpc can be adopted for the H {\sc i} feature.
 At this distance the linear diameter of the H {\sc i} shell is 300 pc, the H {\sc i}  mass (calculated by integrating all contributions within the shell across all the channels where the shell is present) amounts  $1.4 \times 10^{5} (d/17 \mathrm{\,kpc})^2 \mathrm{\,M}_\odot$, and the ambient density is $\sim 0.5 (d/17 \mathrm{\,kpc})^{-1} \mathrm{\,cm}^{-3}$ .
 
In this scenario, HESS J1943+2118 would be a PWN left after a SN explosion of which no radio continuum remnant is observed because the shock front expanded into a medium  with very low ambient density. The weak point of this picture is that an expansion velocity along  the line of sight of about 130 km\,s $^{-1}$ is required in order to reconcile the observed central velocity of the shell ($\sim +54$ km\,s$^{-1}$) with the systemic velocity of HESS J1943+2118  ($\sim -77$ km\,s$^{-1}$ if it is located at $\sim$ 17 kpc).   As the shell diameter does not change significantly in size with velocity within the velocity range where it is detected, and  it is not possible to clearly identify other portions of the shell in different velocity ranges because of confusion with Galactic emission, we do not have arguments to prove that the shell expands at such velocity.

\section{DISCUSSION}
\label{disc}

\subsection{The Blazar Hypothesis} \label{HBL}

The lack of optical/infrared spectroscopic measurements makes it
difficult to unambiguously identify J1943+2118 as a blazar, although
\cite{hess_disc} mention a preliminary infrared spectrum that shows
no obvious emission lines. Based on its spectral energy distribution
(SED), the source could belong to the HBL class that make up the
majority of VHE detected AGNs \citep{hess_disc}. According to
\cite{HBL_Wise}, the infrared magnitudes measured by the Wide-field
Infrared Survey Explorer ({\it WISE}) satellite can be used to
distinguish blazars from other extragalactic radio sources. We
checked the now available {\it WISE} All-Sky Data Release
\citep{Wise}. Based on the {\it WISE} infrared magnitudes of the
2MASS source associated with J1943+2118, the source lies within the
so-called {\it WISE} blazar strip, grouped together with confirmed
HBL sources.

If we assume that the TeV source is an HBL  at a
moderate redshift of z=0.3 (most of the HBLs with known
redshifts detected in the TeV regime have a redshift $\le0.3$ according to 
the currently available data of TeVCat) the observed parameters of the detected 
 point-like source (flux density of 31 mJy, angular size of 15.8 mas), 
 imply  a very low brightness temperature of only $T_\mathrm{B}=7.7 \times 10^7$\,K. 
 Even taking into account a correction for angular broadening of the point source 
 produced by the intervening ionized gas (as discussed in Section \ref{EVNres}), 
 the brightness temperature is still substantially lower than the intrinsic 
 equipartition value estimated for relativistic compact jets ($\sim 5 \times 10^{10}$ K, Readhead 1994).
In the case of HBLs, the gamma-ray emission which is Doppler-boosted
in the jet is produced by inverse-Compton scattering of ambient
low-energy photons. The fundamental parameters of the parsec-scale
jets can be inferred from high-resolution VLBI observations. Even
considering that the jet bulk Lorentz factors determined from radio
data of TeV HBLs are modest \citep[$\Gamma$ $\sim$ $3-4$;][and
references therein]{Piner2010}, and there are indications for larger
jet inclinations with respect to the line of sight
\citep[$\sim$$15-30\degr$;][]{Wu2007}, the measured brightness
temperature of J1943+2118 still seems too low. Our value of $T_{\rm
B}$ $<$ $10^{8}$~K is well below the typical brightness temperatures
found for other extensively studied TeV HBLs
\citep[$10^{9}-10^{10}$~K; e.g.][]{Giroletti2006, Piner2008,
Piner2010}.

There are two possible ways to have higher rest-frame brightness
temperature. One is if the source is more distant, located at a
higher redshift. However, even assuming an extremely large redshift
of $z\approx6$ \citep[where the most distant known radio-loud
quasars are located,][]{high_z}, the brightness temperature would
only be $\sim 4 \times 10^8$\,K. Another possibility to obtain a higher brightness temperature value is, as mentioned before, by assuming that the
intrinsic source size is smaller than the observed, because it is
broadened by interstellar scattering effects in the ionized
turbulent medium of the Milky Way.

If we assume that all the flux density recovered by the EVN
measurement (31\,mJy) originates from the compact core of the HBL,
then its intrinsic size at 1.6\,GHz  cannot be larger  than 1.2\,mas
in order to measure a brightness temperature of at least $10^{10}$\,K.
Consequently, the scattering size at 1.6 GHz must be at least
$15.7$\,mas. Such a value is not unprecedented. For example
\cite{FeyCygnus} studied several extragalactic lines of sight
passing through the Cygnus region of the Milky Way. The largest
scattering size measured (at 1.6 GHz) was $28.2$\,mas for the quasar
B2005+403. Two of the most scatter-broadened extragalactic sources
have much larger scattering disk sizes: $298$\,mas and $\sim$
4000\,mas for B1849+005 \citep{1849scat}, and NGC\,6334B
\citep{NGCscatter}, respectively. However, the interstellar
``clumps'' responsible for these values are identified in all these
cases, and consequently included in the NE2001 model of
\cite{ne2001}. In the case of J1943+2118, no such clump is known (at
least yet). \cite{averbroad} studied a sample of AGNs and found that
the average scattering disk size is 2\,mas (at 1\,GHz); the largest
measured value was $7.8$\,mas (unfortunately, no AGN was chosen
from the longitude region $-60\degr\le l \le 60\degr$, where
J1943+2118 is located). \cite{FeyPhD} examined angular broadening of
radio sources near the Galactic Plane in the longitude range
$20\degr<l<90\degr$ and reported that the observed scattering
decreases rapidly with increasing Galactic longitude, reaching a
minimum at $l\approx 60\degr$. Therefore, we do not expect enhanced
scattering in the direction of J1943+2118.

Our results also showed that the EVN data recover only one-third of
the total flux density of J1943+2118. This is unlike what is usually
observed for VLBI-imaged TeV HBLs, where the total radio emission is
dominated by the compact mas-scale core. For example, the largest
difference between the total (NVSS) and the core (VLBI) flux density
is less than 50 per cent in the sample of \cite{Giroletti2006}.
This, and the measured low brightness temperature are indications
against the beamed blazar hypothesis.

\subsection{The Galactic Gamma-Ray Binary Hypothesis} \label{gamma}

Although \cite{hess_disc} excluded the possibility that
HESS\,J1943+213 is related to a gamma-ray binary, because no bright,
early-type star was detected in the system, we discuss whether the
observed radio properties fit to this scenario.

There are three objects reported in the literature that are
considered as classical gamma-ray binaries, PSR~B1259-63, LS~5039,
and LSI~+61 303. In these systems, the binarity is well established.
They show GeV and/or TeV emission, with a peak in the spectral
energy distribution at MeV--GeV energies. Recently, a fourth system
was discovered, HESS~J0632+057 \citep[][and references
therein]{moldon2011}. In all of these systems, the emission is
variable and periodic in all wavebands, in contrast with our target
HESS~J1943+213. All of the sources mentioned above were detected
with VLBI technique on mas scales, with extended and variable
structure. The observed brightness temperature due to the resolved
structure on mas scales in HESS~J1943+213 would be consistent with
this scenario. With monitoring observations on VLBI scales at
somewhat higher resolution than in our presented exploratory EVN
data, one could probe the variable appearance of the source on mas
scales. (Moreover, a Galactic source is expected to show a proper
motion detectable with VLBI in the future.)

The multi-wavelength properties of gamma-ray binaries have been
interpreted with either microquasar jets \citep[e.g.][]{bosch2009},
or particle acceleration in shocks between a relativistic pulsar
wind and the wind from the companion star \citep[e.g.][]{tavani97,
dubus2006}. Most notably, in the case of LSI~+61 303,
\cite{dhawan2006} showed that the microquasar interpretation of the
system does not agree with the variable structure observed with the
Very Long Baseline Array (VLBA), showing orbital modulation of the
extended structure as expected in the shocked-wind scenario.

What is however not observed in gamma-ray binaries is the prominent,
arcminute-scale radio structure present in HESS~J1943+213. It is
hard to reconcile with either the microquasar jet or the colliding
wind scenario. Therefore, in addition to the missing optical
counterpart and the lack of (periodic) variability, we find another
piece of evidence against the gamma-ray binary interpretation of
HESS~J1943+213.

\subsection{Other Galactic Origin} \label{PWN}

In the following, we investigate the hypothesis whether
HESS\,J1943+213 (and consequently its radio counterpart, J1943+2118)
can be a Galactic object, specifically a PWN.

If the paradigmatic PWNe Crab and 3C 58 were located at a
distance of 17\,kpc, as the proposed distance for J1943+2118, their
angular size in radio wavelengths would be of the order of $\sim
0\farcm8$ for Crab and $\sim 1\farcm7$ for 3C~58. This is in
excellent agreement with the size of the structure around J1943+2118
as seen from the VLA-C observations (Fig. \ref{archVLA}). The size
of the PWN in X-rays is generally smaller than in radio due to the
smaller synchrotron lifetimes of the higher-energy electrons
\citep{crab}, thus explaining why {\it Chandra} detected it as a
point-like source. Besides, the radio spectral index $-0.32$
reported by \cite{vollmer_cat} for the NVSS point source coincident
with J1943+2118 is compatible with the standard spectral indices of
PWNe \citep[between $0$ and $-0.3$, ][]{pwn_spectrum}. In that
sense, the new radio  results  do not contradict to a
PWN hypothesis.

Even when we cannot unambiguously prove that the H {\sc i} shell and HESS\,J943+213 are associated, it is worthwhile to explore the idea of having gamma-ray emission related to peculiar features in the
Galactic gas. \citet{renaud2008} showed that the
``dark'' VHE source HESS\,J1503$-$582, which lacks any traditional
counterpart, is spatially coincident with the Forbidden Velocity
Wing FVW 319.8+0.3, and proposed that the objects can be
associated. \cite{kang2007} reported on the existence of 81 FVWs. The FVWs are faint, wing-like H {\sc i}
  features at velocities beyond the boundaries allowed by Galactic
  rotation. Later \cite*{kang2010} explored some of them at higher
  resolution using the Arecibo and Green Bank radio telescopes, concluding 
  that a significant fraction of the FVWs are expanding
  shells with physical parameters consistent with those expected from
  very old supernova remnants (SNRs).
 
In this context it is interesting to investigate  how often a TeV source can be found
associated with a large H {\sc i} shell and/or FVW structures. As a
first approximation we confronted lists of high-energy sources with the results of \citet{mcclure2002} (and references therein) in their search for large H {\sc i}
shells  in a field of over 2000 square degrees in the third and
fourth Galactic quadrants (region $253^\circ \leq l \leq 358^\circ$,
$-10^\circ \leq b \leq +10^\circ$). We found the following: the
studied region contains 19 Galactic H {\sc i} shells and 31
TeV gamma-ray sources. Among the 31 TeV sources, 11 were identified
with galactic objects (shell SNRs, massive stars clusters and X-ray
binaries), 11 as PWN and 9 still remain unidentified. The
cross-checking between H {\sc i} features and TeV sources indicates
that among the 11 identified with galactic objects, only 1 coincides
with a H {\sc i} shell; from the 11 classified as PWN, 2 appear
related to H {\sc i} shells; and, notably, 7 out of the 9
unidentified TeV sources lie clearly within or on the borders of H
{\sc i} shells (Dubner et al., in preparation). The sizes and
ambient densities of these shells are comparable to those of our
discovery.

Though the study is in progress and the statistic is incomplete,
these numbers are suggestive that the positional agreement shown by
\citet{renaud2008} may be a quite common phenomenon. That is, very
high energy point-like or extended sources which could not be
identified with known Galactic or extragalactic objects might be
related with large H {\sc i} shells, the fossil remains  of
the combined action of powerful stellar winds and supernova
explosions.


\section{SUMMARY}

\label{concl}

The exploratory EVN observation allowed us to pinpoint with high
precision a radio source spatially coincident with  the {\it
Chandra} X-ray source proposed to be the counterpart of the TeV
source  HESS\,J1943+213. It confirms that all the reported
multi-wavelength sources are physically related to the same object.
The EVN observation also revealed that the radio source is highly
resolved, and its brightness temperature is significantly lower than
the intrinsic equipartition value estimated for relativistic compact
jets \citep{equi}. Additionally, from the study of the radio
continuum emission at intermediate angular resolution based on
archival VLA C-array observations, we could identify the presence of
an elongated feature, about 1\arcmin ~in size, around the point
source. These results from radio continuum observations pose
difficulties to the TeV source identification as a HBL. To be able
to reconcile the large source size with the object being a HBL, we
have to introduce angular scattering by a presently unknown
intervening galactic ``clump'' of ionized matter which should be
similar in the scattering characteristics to the ionized material in
the Cygnus super-bubble region, however much smaller in size. This
hypothesis could be investigated via future multi-frequency VLBI
observations, since we expect that at low frequencies (below at
least $\sim 10$\,GHz), the apparent source size would change
according to the scattering law, roughly with the square
of the observing frequency.

We also discussed the possibility of HESS\,J1943+213 being a
gamma-ray binary. While part of the characteristics detected by our
EVN observation can be explained within a gamma-ray binary scenario,
none of the known gamma-ray binaries exhibit large arcminute-scale
structure, similar to the one revealed by the archival VLA data in
HESS\,J1943+213. This finding, together with the missing early-type
companion star and the lack of periodic variability are indications
against the gamma-ray binary hypothesis.

Additionally, we studied the large-scale  H {\sc i} emission in
an extended field around the TeV source. The H {\sc i} spectral line
data revealed a shell-like feature around HESS\,J1943+213. In spite that we
do not yet have an unequivocal evidence confirming that the two
objects are associated, a scenario where the
TeV emission detected by H.E.S.S. is produced by a distant PWN would be compatible with some imprints left in the interstellar medium by the past events that ended in a supernova explosion in the region.
This object would be placed in the outermost parts of the Galaxy,
beyond 16\,kpc according to \citet{leahy2012}, but still within the
external arms of the Milky Way. 

Higher-resolution VLA observations already initiated by our group would be able to shed light on
the structure of the arcminute-scale radio feature and its relation
to the compact object detected by our EVN observation. In the PWN
scenario, if we assume that the compact bright feature detected by
EVN is solely emitted by the pulsar itself, then according to e.g.
\citet{hankins93}, the observed flux density must be beamed and
intrinsically periodic. The scattering along the line of sight to
HESS\,J1943+213, when put at 17\,kpc distance, would not smear out
the pulses \citep{ne2001}. Thus, searching for pulsed radio emission
would help to reveal the nature of the object, but this tool is useful only
 if the pulsar responsible for  blowing the PWN is  beamed towards us.

Finally, from a quick inspection of the distribution of the large H
{\sc i} shells and FVW structures that are being discovered in our
Galaxy, and the distribution of the unidentified HESS sources, we
find that a significant fraction (7 out of 9 in the third and fourth
Galactic quadrants) are positionally coincident. This result is part
of a larger study underway, and at present is only a suggestive
conclusion.

To summarize, based upon the currently available observational
facts, the true nature of the TeV source HESS\,J1943+213 still cannot be
decided unambiguously. However, if we conclude that it is a HBL
object, then it is either not beamed, in contrast with what we
currently know about HBLs, or a significant amount of highly
turbulent ionized material is needed in the line of sight towards
this source to explain its apparent size. On the other hand, if the
source is indeed a PWN, and is related to the discovered H {\sc i}
feature, then the importance of the reported discovery resides not
only in the fact that it naturally explains the origin of the TeV
emission, but also, as remarked by \cite{2006Koo}, it helps to solve
the long-standing problem of the ``missing'' Galactic SNRs, where
the detected SNRs barely make up 1 per cent of the expected number.
It also opens a new path to investigate the nature of the dark TeV
sources.

\acknowledgments

We are grateful to the chair of the EVN Program Committee, Tiziana
Venturi, for granting us short exploratory e-VLBI observing time in
May 2011. We would like to thank Joeri van Leeuwen for useful
discussion and comments. We are grateful to the anonymous referee for suggestions that improved this work significantly. The EVN is a joint facility of European,
Chinese, South African, and other radio astronomy institutes funded
by their national research councils. The NRAO is a facility of the
National Science Foundation operated under cooperative agreement by
Associated Universities, Inc. This work was supported by the
European Community's Seventh Framework Programme, Advanced Radio
Astronomy in Europe, grant agreement no.\ 227290, and Novel
EXplorations Pushing Robust e-VLBI Services (NEXPReS), grant
agreement no.\,RI261525. This research was supported by the
Hungarian Scientific Research Fund (OTKA, grant no.\ K72515 and K104539). GD and
EG are members of CIC-CONICET (Argentina) and their work is
partially supported by CONICET, ANPCYT and UBACYT grants from
Argentina. This publication makes use of data products from the
Wide-field Infrared Survey Explorer, which is a joint project of the
University of California, Los Angeles, and the Jet Propulsion
Laboratory/California Institute of Technology, funded by the
National Aeronautics and Space Administration. We acknowledge the
role of W.W. Tian in leading us to apply more scrutiny when looking
for evidence of the Galactic nature of HESS\,J1943+213.



{\it Facilities:} \facility{EVN}, \facility{VLA}.




\bibliographystyle{apj}
\bibliography{ref}

\clearpage

\end{document}